\documentstyle[aps,prl]{revtex} 
\draft
\begin{document}
\twocolumn[\hsize\textwidth\columnwidth\hsize\csname@twocolumnfalse
\endcsname
]
\noindent
{\large \bf Valley Phase Transition in Si MOSFETs}
\vskip 0.5cm
~~~The recent detailed experiments on high mobility Si
 MOSFETs\cite{krav,lubkin} have convincing supports for
 the existence of a phase transition at low densities
 around a sample-dependent value of
 $n_s=1\times 10^{11}\,\mbox{cm}^{-2}$. This density
 translates to $r_s=8.64$ where $r_s=1/a_B^*\sqrt{\pi n_s}$
 and $a_B^*$ is the effective Bohr radius. The driving
 mechanism of the transition is, however not yet resolved.
 The role of superconductivity in this connection is
 suggested recently by several groups\cite{phil,belitz}.
 Other theoretical works characterized the insulator phase
 as the disorder driven electron solid\cite{tanat} and
 metastable frozen electron solid\cite{thakur}. We would
 like to contribute to this on-going discussion by
 suggesting the possibility of a valley phase transition.

~~~Originally, Bloss, Sham and Vinter\cite{bloss} put
 forward the idea of a valley phase transition in Si(100)
 inversion layers having two degenerate valleys, that
 initiated some theoretical and experimental activity in
 the beginning of 1980's\cite{cole,isih}. The driving
 mechanism of this transition is many-body in origin:
 stacking the available
 inversion layer electrons just to one of the valleys
 (degenerate in single-electron level) increases the total
 kinetic energy as well as the intravalley exchange
 energy. However, the exchange energy is negative and towards
 lower electronic densities, this single valley occupation
 becomes the ground state. Based on the density functional
 calculations, Bloss {\it et al.} predicted the valley phase
 transition density to be around
 $n_s=3\times 10^{11}\,\mbox{cm}^{-2}$. They also conjectured
 the low density phase (i.e.,
 $n_s < 3\times 10^{11}\,\mbox{cm}^{-2}$) to have
 thermally activated conductivity due to formation of spatial
 domains where the populated valley toggles abruptly, as in
 ferromagnetic materials. To verify these theoretical claims,
 Cole {\it et al.}\cite{cole} performed intersubband transition
 experiments on Si(100) MOSFETs, and {\it only for the high
 mobility ones}, they registered a change in the slope of the
 subband transition energy around
 $5\times 10^{11}\,\mbox{cm}^{-2}$ in the direction that complies
 with the valley phase transition explanation. The conductivity
 was observed to be thermally activated for densities lower
 than $2\times 10^{11}\,\mbox{cm}^{-2}$. The following calculations
 of Isihara and Ioriatti\cite{isih} on the ground state energies of
 the 2D electron gas having single and double valley degeneracies
 confirmed that for $r_s>8.011$ single valley occupation becomes
 lower in energy than the equal population. 

~~~These available results in the past literature seem to be
 relevant with the recently observed effects in Si
 MOSFETs\cite{krav,lubkin}. The possibility of a valley phase
 transition can be experimentally checked in these MOSFET
 samples by probing the valley degeneracy on either side of the
 transition density. Furthermore, it will be interesting to see the
 results of the same detailed experiments for the GaAs/AlGaAs HEMTs
 having a valley degeneracy of one; in turn, a phase transition
 ruled by exchange interactions along the same lines can only
 be due to spin degeneracy, leading this time to a
 {\it ferromagnetic} state
 beyond some density. However, regardless of the validity of the
 valley phase transition picture for the driving mechanism, much
 more theoretical
 and experimental labor is required to clarify the observed
 metallic-like and insulator-like phases.
\vskip 0.5cm
\noindent C. Bulutay$^*$ and M. Tomak$^{\dagger}$
\smallskip\par
{\small $^*$Department of Electrical and Electronics Engineering, \par
        \ Middle East Technical University, Ankara 06531, Turkey\par
\smallskip\par
        $^{\dagger}$Department of Physics,  \par
        \ Middle East Technical University, Ankara 06531, Turkey}

\end{document}